\documentstyle[12pt,axodraw,epsf,epsfig]{article}
\advance\textheight by \topskip
\newcommand{\sla}{\kern -5.4pt /}
\newcommand{\Dir}{\kern -6.4pt\Big{/}}
\newcommand{\Dirin}{\kern -10.4pt\Big{/}\kern 4.4pt}
\newcommand{\DDir}{\kern -7.6pt\Big{/}}
\newcommand{\DGir}{\kern -6.0pt\Big{/}}

\newcommand{\ra}{\rightarrow}
\newcommand{\be}{\begin{equation}}
\newcommand{\ee}{\end{equation}}
\newcommand{\bea}{\begin{eqnarray}}
\newcommand{\eea}{\end{eqnarray}}
\newcommand{\beanon}{\begin{eqnarray*}}
\newcommand{\eeanon}{\end{eqnarray*}}
\newcommand{\ba}{\begin{array}}
\newcommand{\ea}{\end{array}}
\newcommand{\bi}{\begin{itemize}}
\newcommand{\ei}{\end{itemize}}
\newcommand{\ben}{\begin{enumerate}}
\newcommand{\een}{\end{enumerate}}
\newcommand{\bc}{\begin{center}}
\newcommand{\ec}{\end{center}}

\newcommand{\bfig}{\begin{center}\begin{picture}}
\newcommand{\efig}[1]{\end{picture}\\{\small #1}\end{center}}
\newcommand{\flin}[2]{\ArrowLine(#1)(#2)}
\newcommand{\wlin}[2]{\DashLine(#1)(#2){2.5}}

\newcommand{\glin}[3]{\Photon(#1)(#2){2}{#3}}

\newcommand{\sof}{\SetOffset}
\newcommand{\bmip}[2]{\begin{minipage}[t]{#1pt}\bfig(#1,#2)}
\newcommand{\emip}[1]{\efig{#1}\end{minipage}}

\newcommand{\ord}{{\cal O}}

\newcommand{\NP}[1]{{\it Nucl.\ Phys.\ }{\bf #1}}
\newcommand{\PL}[1]{{\it Phys.\ Lett.\ }{\bf #1}}
\newcommand{\ZP}[1]{{\it Z.\ Phys.\ }{\bf #1}}

\newcommand{\CPC}[1]{{\it Comp. Phys. Comm.\ }{\bf #1}}

\begin{document}
\tolerance=100000
\thispagestyle{empty}
\setcounter{page}{0}

\begin{flushright}
{\large DFTT 48/97}\\ 
{\rm September 1997\hspace*{.5 truecm}}\\ 
hep-ph/9709454
\end{flushright}

\vspace*{\fill}

\bc     
{\Large \bf QCD corrections to charged current processes
at the Next Linear Collider: CC10, CC11 and CC20 at ${\cal O}(\alpha_s )$. 
\footnote{ Work supported in part by Ministero 
dell' Universit\`a e della Ricerca Scientifica.\\[2 mm]
e-mail: maina@to.infn.it, pittau@psw218.psi.ch, pizzio@to.infn.it}}\\[2.cm]
{\large Ezio Maina$^{a}$, Roberto Pittau$^{b}$ and Marco Pizzio$^{a}$}\\[.3 cm]
{\it $^{a}$Dip. di Fisica Teorica, Universit\`a di Torino
     and INFN, Sezione di Torino,}\\
{\it v. Giuria 1, 10125 Torino, Italy.}\\[5 mm]
{\it$^{b}$Paul Scherrer Institute}\\
{\it CH-5232 Villigen-PSI, Switzerland}

\ec

\vspace*{\fill}

\begin{abstract}
{\normalsize
\noindent
One-loop QCD corrections to the full gauge invariant set of electroweak diagrams
describing the charged current processes 
$e^+ e^- \to u~\bar d~\mu~\bar\nu_\mu$ (CC10),
 $e^+ e^- \to u~\bar d~e^-~\bar\nu_e$ (CC20)
and $e^+ e^- \to u~\bar d~s~\bar c$ (CC11) are computed.
We compare the exact calculation with a ``naive''approach to strong 
radiative corrections which has been widely used in the literature
and discuss the phenomenological relevance of QCD
corrections for NLC physics.
}
\end{abstract}

\vspace*{\fill}
\newpage
\section*{Introduction}
The measurement of $W$ mass to high precision is one of the goals of the NLC
and will provide a stringent test of the Standard Model (SM) \cite{lep2,Wmass}.
In fact, the mass of the $W$ boson in the SM is tightly constrained and 
an indirect determination of $M_W$
can be obtained from a global fit of all present electroweak data.
The fit gives \cite{EWgroupsummer96}
$M_W = 80.338 \pm 0.040^{+0.009}_{-0.018} {\rm GeV}$
where the central value corresponds to $M_H = 300$ GeV and
the second error reflects the change of $M_W$ when the Higgs mass
is varied between 60 and 1000 GeV. 
A disagreement between the value of $M_W$ derived from the global fit and the
value extracted from direct measurement would represent a major failure of the
SM. 
The NLC is expected to reduce the error on the measurement of $W$ mass down to
about 15 MeV \cite{ecfa-desy}, to be compared with an error,
assuming design luminosity,
of about 50 MeV at Lep 2.

In order to extract the desired information from $WW$ production data,
theoretical predictions with uncertainties smaller than those which are foreseen
in the experiments are necessary. This requires a careful study of all radiative
corrections which have to be brought under control. 
In this note we will be concerned with QCD corrections which are known
\cite{MP,CC10} to modify the shape of the $W$--mass peak and the distributions
of others kinematical variables. Since event shape variables are often used to
extract $WW$ production from the background it is highly desirable to have a
complete next--to--leading order (NLO)
study of these distributions for $WW$ events. Furthermore,
calculations of QCD corrections to $\ord (\alpha^2 \alpha_s^2)$ four--jet
production have recently appeared \cite{4jNLO}. Combining these results with a
complete
calculation of QCD corrections to all $\ord (\alpha^4)$ four--fermion processes 
it would be possible to obtain NLO predictions for any four--jet shape
variable at the NLC providing new means of testing perturbative QCD.

In this contribution we present the complete calculation of QCD corrections 
to $e^+ e^- \to u~\bar d~\mu~\bar\nu_\mu$,
 $e^+ e^- \to u~\bar d~e~\bar\nu_e$
and $e^+ e^- \to u~\bar d~s~\bar c$.
.
While this is only a first step in the calculation of all $\ord (\alpha^4)$
four--fermion processes at NLO, these reactions include
the most important source of
background, namely single $W$ production, providing a natural setting for a
first study of the role of QCD corrections to gauge invariant sets of
four--fermion production diagrams.
A gauge invariant description of $e^+ e^- \to u~\bar d~s~\bar c$
requires in the unitary gauge the eleven diagrams shown in fig. 1.
This amplitude is known as CC11 in the literature. The subset of three diagrams
labeled (e) and (f) in fig. 1, in which both $W$'s can go on mass--shell,
is known as CC03 and is often used for quick estimates of $WW$ production.
The ten diagrams required for $e^+ e^- \to u~\bar d~\mu~\bar\nu_\mu$ (CC10) and
the twenty diagrams required for 
$e^+ e^- \to u~\bar d~e~\bar\nu_e$ (CC20) are similar and can be readily
deduced from our figures.
 
In most instances QCD corrections
 have been included ``naively'' with the substitution 
$\Gamma_W \ra \Gamma_W (1 +2/3\,\alpha_s/\pi)$ and multiplying the hadronic
branching ratio by $(1 +\alpha_s/\pi)$. This prescription is exact for CC03 when
fully inclusive quantities are computed. However it can only be taken as 
an order of magnitude estimate even for CC03 in the presence of cuts on jet
properties, as discussed in \cite{MP}
\footnote{ The impact of QCD corrections on the angular distribution of the
decay products of a $W$ and their application to on--shell $W$--pair production
is discussed in ref.\cite{lampe}}.\par
It is well known that differential distributions can be more sensitive to 
higher order corrections than total cross--sections in which virtual and real
contributions tend to cancel to a large degree.
It is therefore necessary to include higher order QCD effects into the
predictions for $WW$ production and decay in a way which allows to impose
realistic cuts on the structure of the observed events.

\bfig(160,245)
\sof(-140,140)
\flin{45,5}{65,25}\flin{65,25}{45,45}
\Text(42,5)[rt]{$e^-$}
\Text(42,45)[rb]{$e^+$}
\glin{65,25}{85,25}{4}
\Text(75,30)[b]{$Z, \gamma$}
\sof(-155,140)
\flin{145,45}{125,65}\flin{125,65}{145,85}
\wlin{113.5,38.5}{125,65}
\flin{120,5}{100,25}\flin{100,25}{120,45}
\Text(149,85)[lb]{$s$}
\Text(149,45)[lt]{$\bar c$}
\Text(124,45)[lb]{$u$}
\Text(124,5)[lt]{$\bar d$}
\Text(90,0)[t]{(a)}
\sof(-10,140)
\flin{45,5}{65,25}\flin{65,25}{45,45}
\Text(42,5)[rt]{$e^-$}
\Text(42,45)[rb]{$e^+$}
\glin{65,25}{85,25}{4}
\Text(75,30)[b]{$Z, \gamma$}
\sof(-25,140)
\flin{145,45}{125,65}\flin{125,65}{145,85}
\wlin{113.5,38.5}{125,65}
\flin{120,5}{100,25}\flin{100,25}{120,45}
\Text(149,85)[lb]{$u$}
\Text(149,45)[lt]{$\bar d$}
\Text(124,45)[lt]{$s$}
\Text(124,5)[lt]{$\bar c$}
\Text(90,0)[t]{(b)}
\sof(120,180)
\flin{45,5}{65,25}\flin{65,25}{45,45}
\Text(42,5)[rt]{$e^-$}
\Text(42,45)[rb]{$e^+$}
\glin{65,25}{85,25}{4}
\Text(75,30)[b]{$Z,\gamma$}
\sof(105,180)
\flin{120,5}{100,25}\flin{100,25}{120,45}
\wlin{113.5,11.5}{125,-15}
\flin{145,-35}{125,-15}\flin{125,-15}{145,5}
\Text(124,45)[lb]{$s$}
\Text(124,5)[lb]{$\bar c$}
\Text(149,5)[lb]{$u$}
\Text(149,-35)[lt]{$\bar d$}
\Text(90,-40)[t]{(c)}
\sof(-140,60)
\flin{45,5}{65,25}\flin{65,25}{45,45}
\Text(42,5)[rt]{$e^-$}
\Text(42,45)[rb]{$e^+$}
\glin{65,25}{85,25}{4}
\Text(75,30)[b]{$Z,\gamma$}
\sof(-155,60)
\flin{120,5}{100,25}\flin{100,25}{120,45}
\wlin{113.5,11.5}{125,-15}
\flin{145,-35}{125,-15}\flin{125,-15}{145,5}
\Text(124,45)[lb]{$u$}
\Text(124,5)[lb]{$\bar d$}
\Text(149,5)[lb]{$s$}
\Text(149,-35)[lt]{$\bar c$}
\Text(90,-40)[t]{(d)}
\sof(-10,40)
\flin{45,5}{65,25}\flin{65,25}{45,45}
\Text(42,5)[rt]{$e^-$}
\Text(42,45)[rb]{$e^+$}
\glin{65,25}{85,25}{4}
\Text(75,30)[b]{$Z,\gamma$}
\wlin{85,25}{97,45}
\wlin{85,25}{97,5}
\flin{117,32}{97,45}\flin{97,45}{117,65}
\flin{117,-15}{97,5}\flin{97,5}{117,18}
\Text(121,65)[lb]{$s$}
\Text(121,32)[l]{$\bar c$}
\Text(121,18)[l]{$u$}
\Text(121,-15)[lt]{$\bar d$}
\Text(75,-20)[t]{(e)}
\sof(120,60)
\flin{65,-15}{65,25}\flin{65,25}{45,45}
\flin{45,-35}{65,-15}
\Text(42,-35)[rt]{$e^-$}
\Text(42,45)[rb]{$e^+$}
\wlin{65,25}{95.5,25}
\wlin{65,-15}{95.5,-15}
\flin{110.5,13}{95.5,25}\flin{95.5,25}{110.5,45}
\flin{110.5,-35}{95.5,-15}\flin{95.5,-15}{110.5,-3}
\Text(114.5,45)[lb]{$u$}
\Text(114.5,13)[l]{$\bar d$}
\Text(114.5,-3)[l]{$s$}
\Text(114.5,-35)[lt]{$\bar c$}
\Text(75,-40)[t]{(f)}
\efig{Figure 1: { Tree level diagrams for 
$e^+ e^- \to u~\bar d~s~\bar c$. The dashed lines are $W$'s.}}

\bfig(160,220)
\sof(-100,170)
\flin{45,0}{25,20}\flin{25,20}{45,40}
\wlin{0,20}{25,20}
\GOval(25,20)(6,6)(0){1}
\Text(25,20)[]{1}
\Text(50,20)[l]{$=$}
\flin{110,0}{90,20}\flin{90,20}{110,40}
\wlin{70,20}{90,20}
\Gluon(105,5)(105,35){-2}{4}
\sof(-100,100)
\flin{45,0}{25,20}\flin{25,20}{45,40}
\glin{0,20}{25,20}{6}
\wlin{25,20}{50,20}
\GOval(25,20)(6,6)(0){1}
\Text(25,20)[]{2}
\Text(63,20)[l]{$=$}
\sof(-100,95)
\glin{85,25}{100,25}{4} \flin{125,0}{100,25} \flin{100,25}{125,50}
\wlin{119,44}{125,63}
\Gluon(110,15)(110,35){-2}{3}
\Text(135,25)[l]{$+$}
\sof(-30,95)
\glin{85,25}{100,25}{4} \flin{125,0}{100,25} \flin{100,25}{125,50}
\wlin{115,40}{125,63}
\Gluon(120,5)(120,45){-2}{5}
\Text(135,25)[l]{$+$}
\sof(40,95)
\glin{85,25}{100,25}{4} \flin{125,0}{100,25} \flin{100,25}{125,50}
\wlin{115,40}{125,63}
\GlueArc(112.5,37.5)(10,-135,45){2}{4}
\Text(135,25)[l]{$+$}
\sof(110,95)
\glin{85,25}{100,25}{4} \flin{125,0}{100,25} \flin{100,25}{125,50}
\wlin{120,45}{125,63}
\GlueArc(110,35)(10,-135,45){2}{4}
\sof(-100,30)
\flin{45,0}{25,20}\flin{25,20}{45,40}
\glin{0,20}{25,20}{6}
\wlin{25,20}{50,20}
\GOval(25,20)(6,6)(0){1}
\Text(25,20)[]{3}
\Text(63,20)[l]{$=$}
\sof(-100,25)
\glin{85,25}{100,25}{4} \flin{125,0}{100,25} \flin{100,25}{125,50}
\wlin{119,6}{125,-13}
\Gluon(110,15)(110,35){-2}{3}
\Text(135,25)[l]{$+$}
\sof(-30,25)
\glin{85,25}{100,25}{4} \flin{125,0}{100,25} \flin{100,25}{125,50}
\wlin{115,10}{125,-13}
\Gluon(120,5)(120,45){-2}{5}
\Text(135,25)[l]{$+$}
\sof(40,25)
\glin{85,25}{100,25}{4} \flin{125,0}{100,25} \flin{100,25}{125,50}
\wlin{115,10}{125,-13}
\GlueArc(112.5,12.5)(10,-45,135){2}{4}
\Text(135,25)[l]{$+$}
\sof(110,25)
\glin{85,25}{100,25}{4} \flin{125,0}{100,25} \flin{100,25}{125,50}
\wlin{120,5}{125,-13}
\GlueArc(110,15)(10,-45,135){2}{4}
\efig{Figure 2: { Basic combinations of loop diagrams. All virtual QCD
corrections to electroweak four fermion processes can be computed using
these three sets. The quark
wave functions corrections are not included because they vanish, in
the massless limit, using dimensional regularization.}} 

\section*{Calculation}
One-loop virtual QCD corrections 
are obtained by dressing all tree diagrams with gluon loops.  
Defining suitable combinations of diagrams as in fig. 2
one can organize all contributions in a very modular way, as
shown in fig. 3. 
As already mentioned in \cite{CC10} all QCD virtual corrections to $\ord
(\alpha^4)$ four--fermion processes can be computed using the set of loop
diagrams shown in fig. 2.

The real emission  contribution 
can be obtained attaching a gluon to the quark line(s) of the tree diagrams 
in all possible positions.
The required matrix elements have been
computed using the formalism presented in ref. \cite{method} with the help of a
set of routines (PHACT) \cite{phact} which generate the building blocks of the
helicity amplitudes  semi-automatically.

The calculation of the virtual corrections has been performed in two different
ways, with identical results. In the first case we have used the standard Passarino--Veltman \cite{PV}
reduction procedure, while in the second we have used the new tecniques
presented in \cite{red}.

In order to be able to integrate separately the real and virtual part 
one has to explicitly cancel all singular contributions
in each term in a consistent way.  
To this aim we have used the subtraction method.
Benefits of this  method are twofold. First, an exact result is obtained and
no approximation needs to be taken; second, all singular terms are canceled 
under the integration sign and not at the end of the calculation,
leading to better numerical accuracies. This is especially relevant for the
present case, since we are aiming for high precision results, with errors
of the order of a per mil.
We have found it particularly convenient to 
implement the dipole formul\ae\, of reference \cite{catani}. These
are a set of completely general factorization expressions 
which interpolate smoothly between the soft eikonal factors and the 
collinear Altarelli--Parisi kernels in a Lorentz covariant way, hence
avoiding any problem of double counting in the region in which partons are both 
soft and collinear. 

All integrations have been carried out using the Monte Carlo routine VEGAS 
\cite{vegas}.

In the absence of a calculation of all 
${\cal O}(\alpha )$ corrections to four--fermion processes, we have only
included the leading logarithmic part of ISR using 
the $\beta$ prescription in the structure functions, where
$\beta = \ln(s/m^2) - 1$. 
Beamstrahlung effects have been ignored.

\section*{Results}
In this section we present a number of cross sections and of distributions at
$\sqrt{s} = 500$ GeV.
We have used $\alpha_s = .123$ in order to conform to the choice made for the
Joint ECFA/DESY Study.
Initial state radiation is included in all results.

For the NLC workshop 
the so called NLC/TH set of cuts have been agreed on:
\bi
\item{} the energy of a jet must be greater than 3 GeV;
\item{} two jets are resolved if their invariant mass is larger than 10 GeV;
\item{} jets can be detected if they make an angle of at least $5^\circ$ with
 either beam.
\ei

\bfig(160,240)
\sof(-140,135)
\flin{45,5}{65,25}\flin{65,25}{45,45}
\Text(42,5)[rt]{$e^-$}
\Text(42,45)[rb]{$e^+$}
\glin{65,25}{85,25}{4}
\Text(75,32)[b]{$Z, \gamma$}
\sof(-155,135)
\flin{145,45}{125,65}\flin{125,65}{145,85}
\wlin{98,25}{125,65}
\flin{120,5}{100,25}\flin{100,25}{120,45}
\Text(149,85)[lb]{$s$}
\Text(149,45)[lt]{$\bar c$}
\Text(124,45)[lb]{$u$}
\Text(124,5)[lt]{$\bar d$}
\GOval(100,25)(6,6)(0){1}
\Text(100,25)[]{2}
\sof(-10,135)
\flin{45,5}{65,25}\flin{65,25}{45,45}
\Text(42,5)[rt]{$e^-$}
\Text(42,45)[rb]{$e^+$}
\glin{65,25}{85,25}{4}
\Text(75,30)[b]{$Z, \gamma$}
\sof(-25,135)
\flin{145,45}{125,65}\flin{125,65}{145,85}
\wlin{113.5,38.5}{125,65}
\flin{120,5}{100,25}\flin{100,25}{120,45}
\Text(149,85)[lb]{$u$}
\Text(149,45)[lt]{$\bar d$}
\Text(124,45)[lt]{$s$}
\Text(124,5)[lt]{$\bar c$}
\GOval(125,65)(6,6)(0){1}
\Text(125,65)[]{1}
\sof(120,175)
\flin{45,5}{65,25}\flin{65,25}{45,45}
\Text(42,5)[rt]{$e^-$}
\Text(42,45)[rb]{$e^+$}
\glin{65,25}{85,25}{4}
\Text(75,30)[b]{$Z,\gamma$}
\sof(105,175)
\flin{120,5}{100,25}\flin{100,25}{120,45}
\wlin{113.5,11.5}{125,-15}
\flin{145,-35}{125,-15}\flin{125,-15}{145,5}
\Text(124,45)[lb]{$s$}
\Text(124,5)[lb]{$\bar c$}
\Text(149,5)[lb]{$u$}
\Text(149,-35)[lt]{$\bar d$}
\GOval(125,-15)(6,6)(0){1}
\Text(125,-15)[]{1}
\sof(-140,55)
\flin{45,5}{65,25}\flin{65,25}{45,45}
\Text(42,5)[rt]{$e^-$}
\Text(42,45)[rb]{$e^+$}
\glin{65,25}{85,25}{4}
\Text(75,32)[b]{$Z,\gamma$}
\sof(-155,55)
\flin{120,5}{100,25}\flin{100,25}{120,45}
\wlin{97,25}{125,-15}
\flin{145,-35}{125,-15}\flin{125,-15}{145,5}
\Text(124,45)[lb]{$u$}
\Text(124,5)[lb]{$\bar d$}
\Text(149,5)[lb]{$s$}
\Text(149,-35)[lt]{$\bar c$}
\GOval(100,25)(6,6)(0){1}
\Text(100,25)[]{3}
\sof(-10,35)
\flin{45,5}{65,25}\flin{65,25}{45,45}
\Text(42,5)[rt]{$e^-$}
\Text(42,45)[rb]{$e^+$}
\glin{65,25}{85,25}{4}
\Text(75,30)[b]{$Z,\gamma$}
\wlin{85,25}{97,45}
\wlin{85,25}{97,5}
\flin{117,32}{97,45}\flin{97,45}{117,65}
\flin{117,-15}{97,5}\flin{97,5}{117,18}
\Text(121,65)[lb]{$s$}
\Text(121,32)[l]{$\bar c$}
\Text(121,18)[l]{$u$}
\Text(121,-15)[lt]{$\bar d$}
\GOval(97,5)(6,6)(0){1}
\Text(97,5)[]{1}
\sof(120,55)
\flin{65,-15}{65,25}\flin{65,25}{45,45}
\flin{45,-35}{65,-15}
\Text(42,-35)[rt]{$e^-$}
\Text(42,45)[rb]{$e^+$}
\wlin{65,25}{95.5,25}
\wlin{65,-15}{95.5,-15}
\flin{110.5,13}{95.5,25}\flin{95.5,25}{110.5,45}
\flin{110.5,-35}{95.5,-15}\flin{95.5,-15}{110.5,-3}
\Text(114.5,45)[lb]{$u$}
\Text(114.5,13)[l]{$\bar d$}
\Text(114.5,-3)[l]{$s$}
\Text(114.5,-35)[lt]{$\bar c$}
\GOval(95.5,25)(6,6)(0){1}
\Text(95.5,25)[]{1}
\efig{Figure 3: One loop gluonic corrections to the $u~\bar d$ line of the
$e^+ e^- \to u~\bar d~s~\bar c$ process. Similar corrections on the $s~\bar c$
line must be included.}

This set of cuts will be referred to as ``canonical'' in the
following.
We have preferred a different criterion for defining jets which is
closer to the actual practice of the experimental collaborations. For mass
reconstruction studies we have used the Durham \cite{durham} scheme 
with $y_D = 1.\times 10^{-3}$.
The four--momenta of the particles which have to be recombined have been simply
summed.
If any surviving jet had an energy smaller than 3 GeV
it was merged with the jet closest in the Durham metric. 

Previous studies \cite{WMC} have shown that the differences between the total 
cross sections obtained from CC10 and CC11 and those obtained with CC03 are at
the per mil level. The non resonant background is far more important when there
is an electron in the final state: the cross section, with canonical cuts,
for CC20 at $\sqrt{s} = 190$ GeV is larger than the cross section calculated
from CC03 by several percent.
When the full NLO results is compared with the naive--QCD (nQCD)predictions,
large effects have been found in observables like the average shift of the
mass reconstructed from the decay products from the true $W$ mass for fully
hadronic processes. 

In fig. 4 and 5 we show some representative distribution for the semileptonic
events. 
In fig. 4a we present the normalized distribution of the angle between
the muon and the closest jet for CC10. 
The events in fig. 4a pass all NLC/TH cut with 
the exception of the minimum $\theta_{\mu j}$. With increasing collider energy,
the two $W$'s tend to fly
apart with larger relative momentum and therefore the probability of a jet to
overlap with the lepton decreases. This behaviour is clearly visible in fig. 4a
which shows that jets are typically well separated 
in angle
from the charged lepton and that the differences between the exact distribution
and the one with ``naive'' corrections is confined to very large angles.

Fig. 4b shows the distribution of the minimum angle between any jet
and either beam. We have separated the contribution
of two--jet and three--jet final
states making it possible to estimate the effect of different angular cuts
on the cross section.

The total NLO cross section with canonical cuts is .23554(4) $pb$,
to be compared with the nQCD results of .23065(3). The two predictions differ by
more than 2\%.

Fig. 5a and 5b present the normalized distributions of the angle between
the electron and the closest jet and of the minimum angle between any jet
and either beam, respectively, for $e^+ e^- \to u~\bar d~e~\bar\nu_e$.

In fig. 6 we compare the NLO spectrum of the average reconstructed $W$ mass 
for CC11 with the nQCD case result.
All events with at least four observed jets have been retained in the plots.
The two candidate masses are obtained forcing all events to four jets, merging
the two partons which are closest in
the Durham scheme, and then selecting the two pairs which minimize
\be\label{selection}
\Delta^\prime_M = \left( M_{R1} 
- M_W \right)^2 + \left( M_{R2} -M_W \right)^2.
\ee
where $M_{R1}\ M_{R2}$ are the two candidate reconstructed masses and $M_W$ is
the input $W$ mass.
In fig. 6 the dashed line refers to the nQCD results while those of the
full NLO calculation are given by the continous line.
Our plot is obtained using only the basic set of cuts described above.

Fig. 6 shows that at NLO the mass distribution is shifted towards lower
masses and a long tail for rather small average masses is generated. Because of 
the larger relative momentum of the two $W$'s it is less likely that partons
from
the decay of one $W$ end up close to the decay products of the other $W$--boson,
therefore the effect is smaller than at Lep 2 energies \cite{CC11}.

In fig. 7 we present the distributions of the
following four--jet shape variables \cite{shape}:
\bi
\item{} the Bengtsson--Zerwas angle: 
 $\chi_{BZ} = \angle [({\bf p_1 \times p_2}),({\bf p_3 \times p_4})]$ (fig. 7a);
\item{} the K\"orner--Schierholz--Willrodt angle :\hfil\break
$\Phi_{KSW} = 1/2 \{\angle [({\bf p_1 \times p_4}),({\bf p_2 \times p_3})]
 + \angle [({\bf p_1 \times p_3}),({\bf p_2 \times p_4})]\}$ (fig. 7b);
\item{} the angle between the two least energetic jets;
$\alpha_{34} = \angle [{\bf p_3},{\bf p_4}]$ (fig. 7c);
\item{} the (modified) Nachtmann--Reiter angle:
$\theta^\ast_{NR} = \angle [({\bf p_1 - p_2}),({\bf p_3 - p_4})]$ (fig. 7d).
\ei
The numbering $i = 1 \dots , 4$ of the jet momenta ${\bf p}_i$ corresponds to
energy--ordered four--jet configurations $( E_1 > E_2 > E_3 > E_4 )$.
We compare the exact NLO results with the distributions
obtained in nQCD and with the results obtained at tree level from the
standard background reactions  $e^+ e^- \to q~\bar q~g~g$ and
$e^+ e^- \to q_1~\bar {q_1}~q_2~\bar {q_2}$.
 
In all subplots of fig. 7 the full NLO results are given by the continous line
and the nQCD prediction is given by the dashed line. The $q~\bar q~g~g$
and $ q_1~\bar {q_1}~q_2~\bar {q_2}$ tree level background distributions are
given by the chain--dotted and the dotted line respectively. 
The shape variables are computed following the procedure outlined in ref.
\cite{aleph} where the Durham cluster algorithm is complemented by the E0
recombination scheme, namely if the two particles $i$ and $j$ are merged the
pseudo--particle which takes their place remains massless, with four--momentum:
\be
E_{new} = E_i + E_j, \hspace{2 cm} 
{\bf p}_{new}=\frac{E_i + E_j}{|{\bf p}_i +{\bf p}_j |} ({\bf p}_i +{\bf p}_j).
\ee
Using this clustering procedure
all five--jet events are converted into four--jet events,
then each event is used in the analysis if $\min _{i,j=1,4} y_{ij} > y_{cut}$
with $y_{cut} = 0.001$. In particular no minimum angle between jets and either
beams is required.

It should be stressed that four--jet shape variables in $WW$ events measure the
correlations between the hadronic decays of the two $W$'s and therefore it
should be explicitely checked whether existing codes, NLO calculations or parton
shower Monte Carlo programs, successfully reproduce the experimental curves.

A separation based on shape--variables of $WW$ events from the background seems,
at first sight, to be rather difficult at the NLC. The variable which most
clearly discriminates between signal and background is the angle between the
two least energetic jets $\alpha_{34}$.
While the signal distribution peaks in the backward direction
the background is almost flat.
On the contrary,
the Bengtsson--Zerwas angle and  the K\"orner--Schierholz--Willrodt angle 
distribution from CC11 are almost indistinguishable from those generated
from the $q~\bar q~g~g$ background.
Some additional sensitivity is provided by the Nachtmann--Reiter angle
distribution.
Both signal and background peak at small angles, but while the former is almost
negligible in the backward direction, the
latter shows a large tail which extends to $180^\circ$. 
The distributions obtained in nQCD are very similar to the full NLO results,
contrary to what happens at lower energies \cite{CC11}. 

\section*{Conclusions}
We have described the complete calculation of QCD radiative corrections 
to the charged current processes $e^+ e^- \to u~\bar d~\mu~\bar\nu_\mu$,
$e^+ e^- \to u~\bar d~e~\bar\nu_e$ and $e^+ e^- \to u~\bar d~s~\bar c$
which are
essential in order to obtain theoretical predictions for $W$--pair production
with per mil accuracy. The amplitudes we have derived are completely
differential, and realistic cuts can be imposed on the parton level
structure of the observed events.
For CC11 we
have presented the distribution of the average reconstructed $W$ mass in
fully hadronic events
and the distribution of several four--jet shape variables.
For CC10 and CC20 the distributions obtained in nQCD are very similar to the
exact NLO results, however the corresponding 
total cross sections can differ by more than 2\%.

\vfill\eject

\newpage
\thispagestyle{empty}
\centerline{
\epsfig{figure=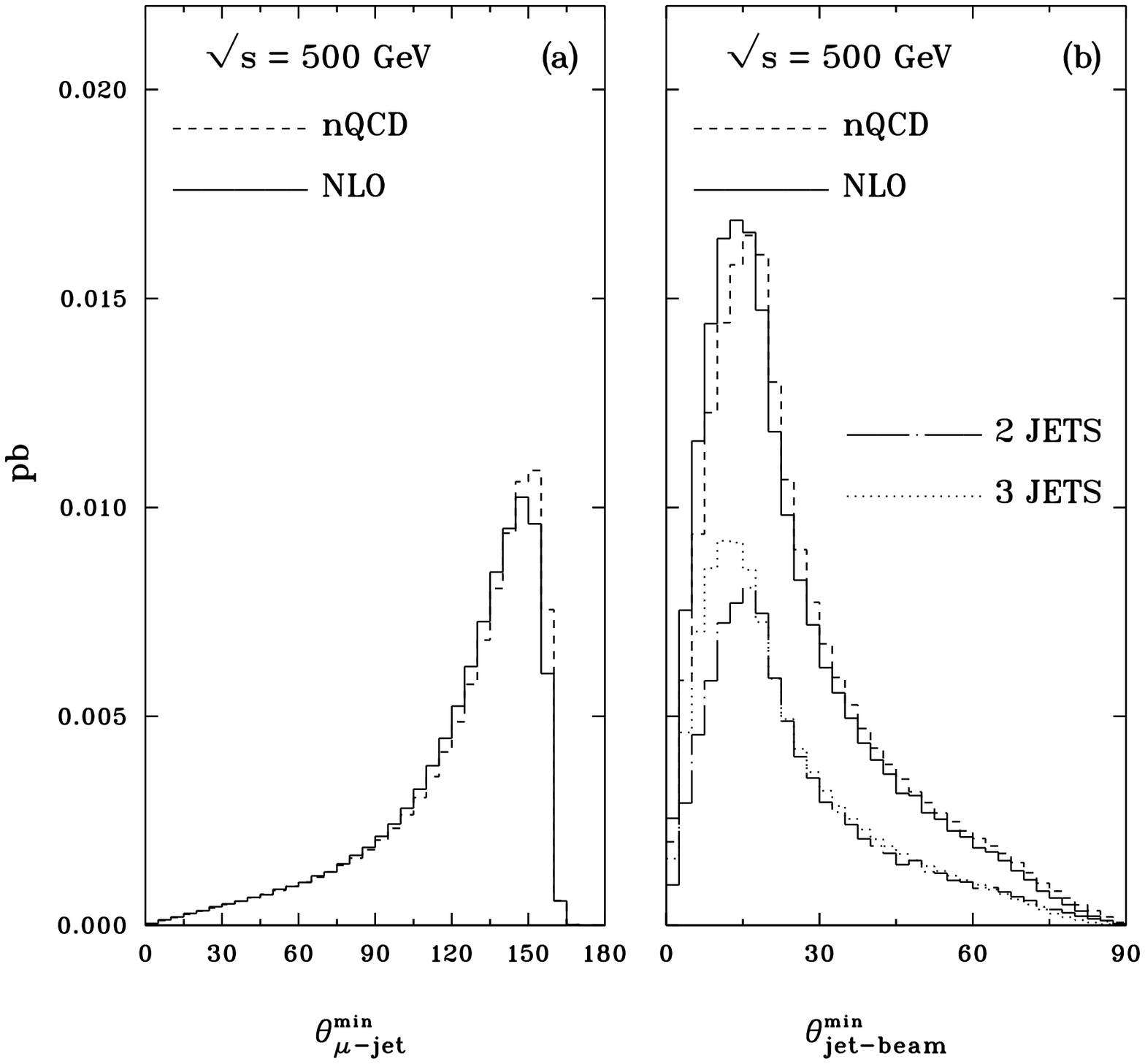,width=15 cm}
}
\vspace*{-3.5 cm}
\noindent
Fig. 4: Distribution of the angular separation of the $\mu^-$ from
the closest jet (a) and of the minimum angular separation between any
jet and either beam (b) at $\sqrt{s} = 500$ GeV for CC10.
The continuous, dotted and dot--dashed histograms are exact NLO results
while the dashed histogram refers to nQCD.

\newpage
\thispagestyle{empty}
\centerline{
\epsfig{figure=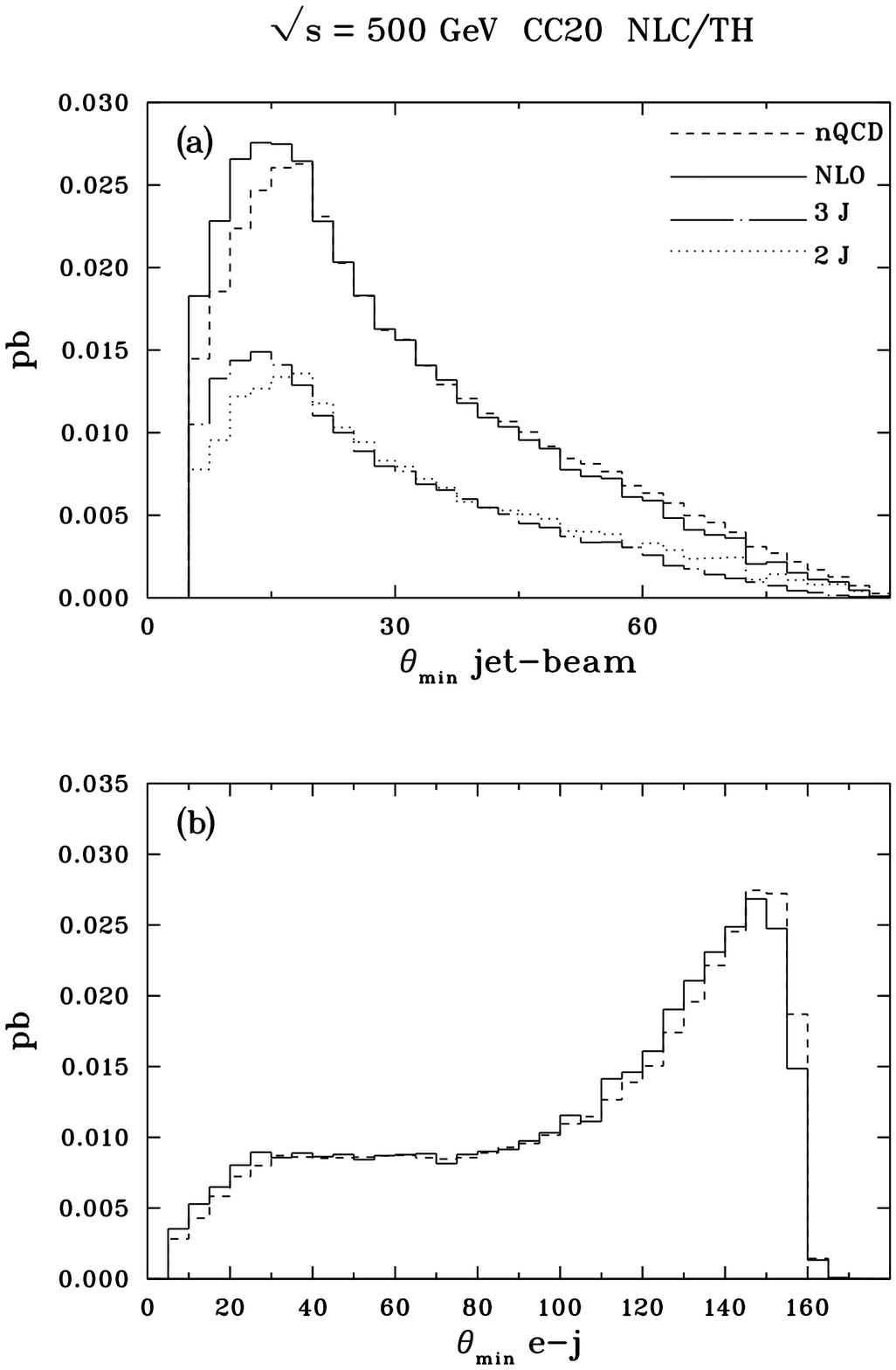,width=16cm}
}
\vspace{-1.9 cm}
\noindent
Fig. 5: Angular separation of the $e^-$ from
the closest jet (a) and between any
jet and either beam (b) at $\sqrt{s} = 500$ GeV.
The continuous, dotted and dot--dashed histograms are exact NLO results
while the dashed histogram refers to nQCD.

\newpage
\thispagestyle{empty}
\centerline{
\epsfig{figure=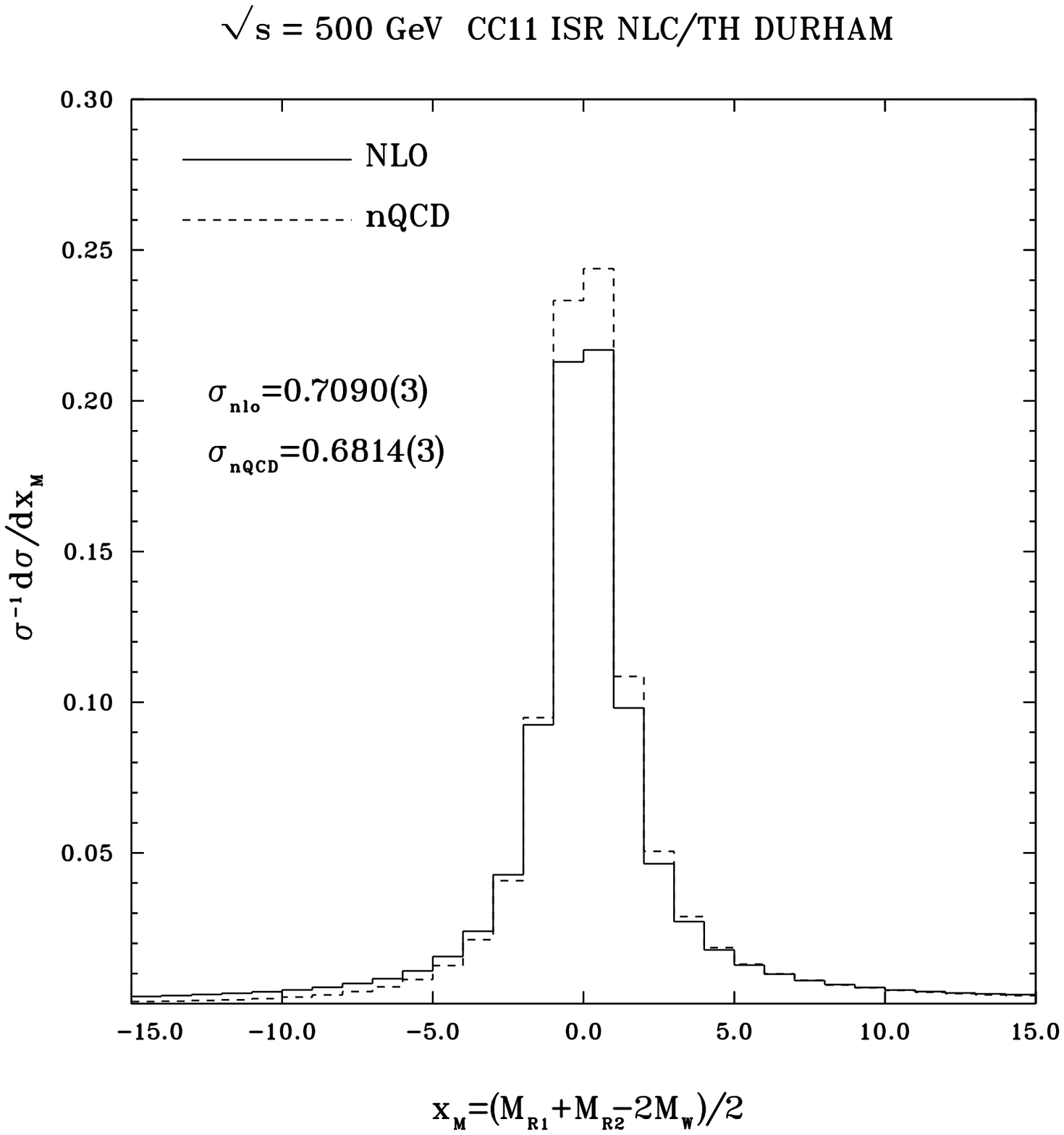,height=21cm}
}
\noindent
Fig. 6: Average mass distribution  at $\sqrt{s} =  500$ GeV.
All NLC/TH cuts are applied.
The continuous histogram is the exact NLO result while the dashed histogram 
refers to nQCD.

\newpage
\thispagestyle{empty}
\centerline{
\epsfig{figure=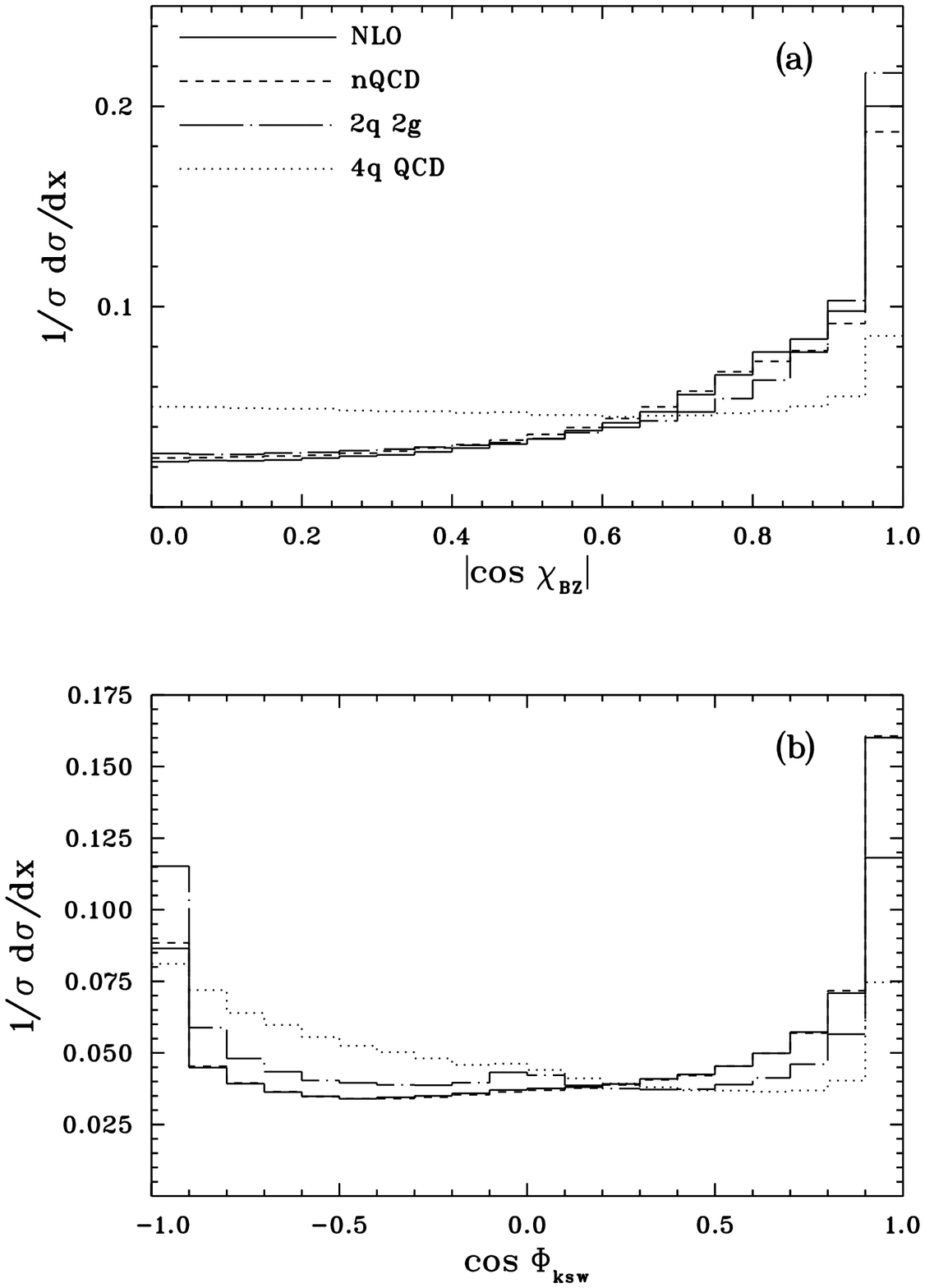,width=16cm}
}
\vspace{-1.9 cm}
\noindent
Fig. 7: Four--jet shape variables at $\sqrt{s} =  500$ GeV.
The full NLO results (continous line) is compared with
the nQCD prediction (dashed line) and with the tree level background
distributions from $q~\bar q~g~g$ (chain--dotted line)
and $ q_1~\bar {q_1}~q_2~\bar {q_2}$ (dotted line).

\newpage
\thispagestyle{empty}
\centerline{
\epsfig{figure=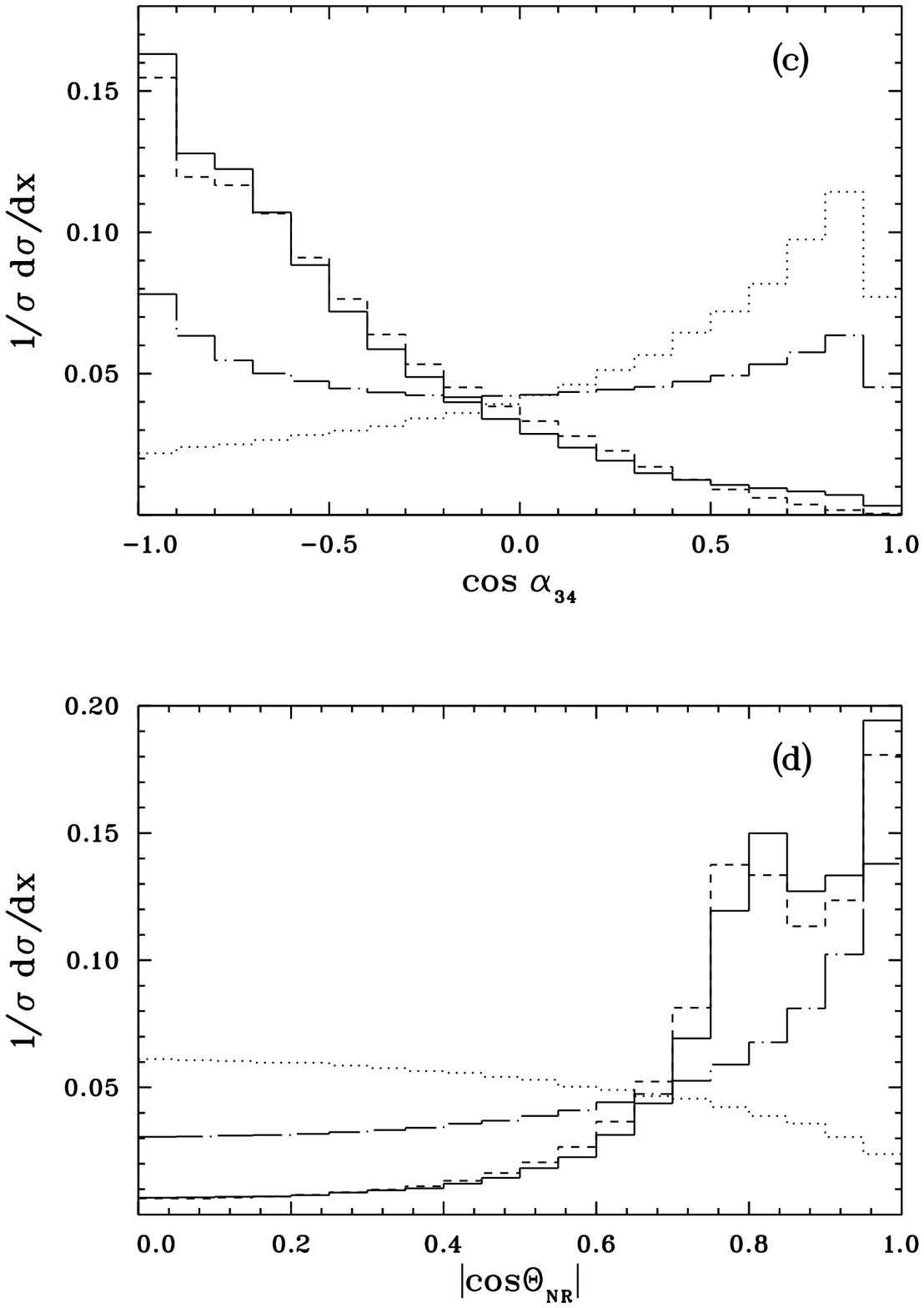,width=16cm}
}
\vspace{-1.9 cm}
\noindent
Fig. 7: Four--jet shape variables at $\sqrt{s} =  500$ GeV, continued.

\end{document}